# Effect of Interlayer Spin-Flip Tunneling for Interlayer Magnetoresistance in Multilayer Massless Dirac Fermion Systems

Kenji Kubo* and Takao Morinari

*Graduate School of Human and Environmental Studies, Kyoto University, Kyoto 606-8501, Japan*

We investigate the effect of the interlayer spin-flip tunneling for the interlayer magnetoresistance under magnetic fields in $\alpha$-(BEDT-TTF)$_2$I$_3$, which is a multilayer massless Dirac fermion system under pressure. The mean field of the spin-flip correlation associated with the interlayer Coulomb interaction enables the interlayer spin-flip tunneling. Assuming the non-vertical interlayer spin-flip tunneling, we calculate the interlayer magnetoresistance using the Kubo formula. The crossover magnetic field, at which the interlayer magnetoresistance changes from positive to negative is shifted by the Zeeman energy and in good agreement with the experiment.

An organic conductor $\alpha$-(BEDT-TTF)$_2$I$_3$ [BEDT-TTF=bis(ethylenedithio)tetrathiafulvalene] is the first bulk massless Dirac fermion system, in which the energy dispersion is linear and the conduction electrons obey the Dirac equation. Below 135K, $\alpha$-(BEDT-TTF)$_2$I$_3$ behaves as an insulator under ambient pressure, where the charge ordering with stripe patterns have been formed[1–3]. For pressures higher than 1.5 GPa, the charge ordering transition is suppressed and the metallic phase is realized at low temperature[4], where the resistivity is almost temperature independent. Meanwhile, the Hall effect is very sensitive to the magnetic field, which means the system is clean.[4] These experimental results shows that the system has extremely narrow energy gap.

Kobayashi and coworkers calculated the energy dispersion,[5,6] which revealed that the bottom of the conduction band contacts with the top of the valence band at the two inequivalent points, called Dirac points, in the first Brillouin zone. These energy dispersion structure is called Dirac cone. Unlike the graphene[7], where Dirac points are in high symmetrical points in the Brillouin zone (see, for a review [8]), the positions of Dirac points in $\alpha$-(BEDT-TTF)$_2$I$_3$ are not at symmetrical points. Moreover, the Dirac cone in $\alpha$-(BEDT-TTF)$_2$I$_3$ is anisotropic and tilted, which is greatly different from the isotropic Dirac cone in graphene. The low energy behavior of the Dirac electrons resides in two valleys near the Dirac points in reciprocal space. Corresponding to these two inequivalent Dirac points, the valley degrees of freedom exist.

α-(BEDT-TTF)$_2$I$_3$ has the layered structure, consisting of the conduction layers of BEDT-TTF molecules and the insulating layers of iodine anions. This multilayer structure is the one of the most important differences, from the single layer massless Dirac fermion system, graphene. We reported the unique spin-ordered states in α-(BEDT-TTF)$_2$I$_3$,[9] where the interlayer antiferromagnetic interaction leads to interlayer ferrimagnetic state in weak magnetic field regime. The characteristic transport phenomena of α-(BEDT-TTF)$_2$I$_3$ are clearly seen in the negative interlayer magnetoresistance under perpendicular magnetic field.[10] Osada gave the formula for the interlayer magnetoresistance[11], which is based on the zero mode Landau level. For relatively large magnetic field, this qualitatively describes the experimental result. On the other hand, the crossover from positive to negative magnetoresistance is observed at low magnetic field. Morinari and Tohyama revealed that this crossover is caused by the non-vertical interlayer tunneling[12]. The non-vertical interlayer tunneling leads to non-vanishing matrix elements between different Landau level wave functions. When the energy gap between nearest neighbor Landau levels is less than the energy scale of the characteristic Landau level broadening, the inter-Landau level mixing on interlayer tunneling leads to the positive magnetoresistance. Although this suggests that the peak temperature of the interlayer magnetoresistance is proportional to $C\sqrt{B}$, where $C$ is a constant and $B$ is a magnetic field, the experiment of the interlayer magnetresistance implies $C\sqrt{B} - g\mu_B B$, where $g$ is the g factor and $\mu_B$ is the Bohr magneton, which is obviously shifted by the Zeeman energy[13]. Konoike and coworkers also suggested the energy shift in the observation of the giant Nernst effect[14]. Therefore, the interlayer tunneling should be related to spin-flip processes.

In this work, we investigate that the effect of the interlayer spin-flip tunneling for the peak temperature of the interlayer magnetoresistance. At the interlayer ferrimagnatic state, the opposite spin term of the Coulomb interaction in the mean field approximation describes the spin-flip interlayer tunneling. Since the opposite spin mean field is finite at low temperature, the spin-flip interlayer tunneling is possible and the peak temperature of the interlayer magnetoresistance is shifted by the Zeeman energy.

We consider the multilayer massless Dirac fermion systems under magnetic fields. We take the xy plane in the 2D layer and the z-direction perpendicular to it. We assume that the Dirac fermion system is realized in each layer. We assume that the valley degrees of freedoms are degenerate and we do not consider the possibility of lifting valley degeneracy. We take the plane wave form with the wave number $k$ in $y$-direction, the Landau level wave functions for Dirac fermions are given by

$$\psi_{n,k}(x,y) = \frac{1}{\sqrt{l_B L_y}} \exp(iky) \Phi_{n,k}(x) \qquad (1)$$

where $l_B$ is the magnetic length defined by $l_B = \sqrt{\hbar/(eB)}$ and $L_y$ is the system dimension in the $y$-direction. The $n$-th Landau level wave function $\Phi_{n,k}(x)$ is written as

$$\Phi_{n,k}(x) = C_n \left[ \begin{pmatrix} -i\,\mathrm{sgn}(n) \\ 0 \end{pmatrix} h_{|n|-1}\left(\frac{x}{l_B} + kl_B\right) + \begin{pmatrix} 0 \\ 1 \end{pmatrix} h_{|n|}\left(\frac{x}{l_B} + kl_B\right) \right]. \qquad (2)$$

Here, $h_{|n|}(\xi)$ is the harmonic oscillator wave function with the eigen value $|n|+1/2$ and the normalization constant $C_n$ is $C_0 = 1$ and $C_n = 1/\sqrt{2}$ for $n \neq 0$. $\mathrm{sgn}(n)$ is the sign of $n$ for $n \neq 0$ and zero for $n = 0$. The field operator represented by this Landau level wave function is given by

$$\hat{\psi}_{l,\sigma}(x,y) = \sum_{n,k} \psi_{n,k}(x,y) c_{l,n,k,\sigma}, \qquad (3)$$

where the $c_{l,n,k,\sigma}$ is the annihilation operator of the $n$-th Landau level with the wave number $k$ and the spin $\sigma$ in the $l$-th layer.

We start from the model which includes the interlayer transfer integral $t_{c'}$ and the interlayer Coulomb interaction, $V_{l,l+1}(c) = e^2/4\pi\varepsilon\sqrt{a_z^2 + c^2}$, where $-e$ is the electron charge, $\varepsilon = 190$ F/m [15] is the dielectric constant and $a_z = 17.5$ Å [16] is the interlayer distance. The non-vertical interlayer tunneling Hamiltonian [12] with the Coulomb interaction is given by

$$\hat{H}_t = -t_{c'} \int d\mathbf{r} \sum_{l,\sigma} \left\{ \hat{\psi}^\dagger_{l,\sigma}(x,y) \hat{\psi}_{l+1,\sigma}(x,y+c'_y) + h.c. \right\} + \sum_{l,\sigma} V_{l,l+1}(c'_y) \int d\mathbf{r}\, \hat{\rho}_{l,\sigma}(x,y) \hat{\rho}_{l+1,\sigma}(x,y+c'_y) \qquad (4)$$

Here, $c'_y$ is a parameter for the non-vertical interlayer tunneling. The density operator is written as

$$\hat{\rho}_{l,\sigma}(x,y) = \hat{\psi}^\dagger_{l,\sigma}(x,y) \hat{\psi}_{l,\sigma}(x,y). \qquad (5)$$

In order to focus on the interlayer spin-flip tunneling, we do not consider the $n$ and $k$ dependence of the wave functions. We include the effect of the spin-ordered state by introducing a potential associated with the mean fields of the spin state. The spin-polarized state is realized by the intralayer ferromagnetic interaction in each layer, while the interlayer ferrimag-netic state is possible by the interlayer antiferromagnetic interaction[9]. The interlayer ferromagnetic spin configuration produces a uniform potential. So, we may focus on the antiferromagnetic component and introduce the staggered magnetization potential $(-1)^l M_{st}$. The Hamiltonian is given by

$$\hat{H} = \left\{ -t_{c'} \sum_{l,\sigma} \left( c^\dagger_{l,\sigma} c_{l+1,\sigma} + h.c. \right) + V \sum_{l,\sigma,\sigma'} c^\dagger_{l,\sigma} c_{l,\sigma} c^\dagger_{l+1,\sigma'} c_{l+1,\sigma'} \right\} + \sum_l (-1)^l M_{st} \left( c^\dagger_{l\alpha} \sigma^z_{\alpha\beta} c_{l\beta} \right). \quad (6)$$

Here, $\sigma^z_{\alpha\beta}$ is the Pauli matrix and $V = V_{l,l+1}(c'_y)$ is the interlayer Coulomb interaction. The labels $n$ and $k$ are implicit in the creation and annihilation operators since we do not consider the $n$ and $k$ dependence now. We apply the mean field approximation to the interlayer Coulomb interaction. The Fourier transform of the mean field Hamiltonian is written as

$$\hat{H}_{MF} = \sum_{q \in RBZ} \begin{pmatrix} c^\dagger_{q\uparrow} & c^\dagger_{q\downarrow} & c^\dagger_{q+Q\uparrow} & c^\dagger_{q+Q\downarrow} \end{pmatrix}$$

$$\times \begin{pmatrix} \varepsilon_q - V X^q_{\uparrow\uparrow,\uparrow\uparrow} & -V X^q_{\downarrow\uparrow,\uparrow\downarrow} & M_{st} & 0 \\ -V X^q_{\uparrow\downarrow,\downarrow\uparrow} & \varepsilon_q - V X^q_{\downarrow\downarrow,\downarrow\downarrow} & 0 & -M_{st} \\ M_{st} & 0 & \varepsilon_{q+Q} + V X^q_{\uparrow\uparrow,\uparrow\uparrow} & V X^q_{\downarrow\uparrow,\uparrow\downarrow} \\ 0 & -M_{st} & V X^q_{\uparrow\downarrow,\downarrow\uparrow} & \varepsilon_{q+Q} + V X^q_{\downarrow\downarrow,\downarrow\downarrow} \end{pmatrix} \begin{pmatrix} c_{q\uparrow} \\ c_{q\downarrow} \\ c_{q+Q\uparrow} \\ c_{q+Q\downarrow} \end{pmatrix}, \quad (7)$$

where $\varepsilon_q = -2t_{c'} \cos(q)$ and $Q = \pi$. Here, we assume a coherent interlayer hopping. This is just for simplifying the calculation of the mean fields $X^q_{\alpha\beta,\gamma\delta}$. The summation is taken over wave numbers in the reduced Brillouin zone. The mean field $X^q_{\alpha\beta,\gamma\delta}$ is defined by

$$X^q_{\alpha\beta,\gamma\delta} = e^{-iq} \frac{1}{N} \sum_{q' \in RBZ} e^{iq'} \left( \langle c^\dagger_{q'\alpha} c_{q'\beta} \rangle - \langle c^\dagger_{q'+Q\alpha} c_{q'+Q\beta} \rangle \right) + e^{iq} \frac{1}{N} \sum_{q' \in RBZ} e^{-iq'} \left( \langle c^\dagger_{q'\delta} c_{q'\gamma} \rangle - \langle c^\dagger_{q'+Q\delta} c_{q'+Q\gamma} \rangle \right). \quad (8)$$

We assume the diagonal elements $\varepsilon_q - VX^q_{\alpha\alpha,\alpha\alpha}$ are renormalized to $\varepsilon'_q = -2t'_{c'}\cos(q)$ with the renormalized transfer integral $t'_{c'}=1$ K. We numerically solve the self-consistent equation about the mean field with the opposite spins $\chi_{\uparrow\downarrow} = \langle c^\dagger_{q\uparrow} c_{q\downarrow}\rangle$.

The result with $M_{st}=0.05$ is shown in Fig. 1. At low temperature, $\chi_{\uparrow\downarrow}$ is finite. The critical temperature depends on the strength of the interlayer Coulomb interaction. In the presence of the finite $\chi_{\uparrow\downarrow}$, the interlayer spin-flip tunneling is possible. When the interlayer Coulomb interaction is larger than the onsite Coulomb interaction, the dimerization state is realized. In this system, the onsite Coulomb interaction is much larger than the interlayer Coulomb interaction. Therefore, the possibility of the dimerized state is ruled out. In the absence of the dimerization state, the interlayer Coulomb interaction gives rise to the interlayer spin correlation, $\chi_{\uparrow\downarrow}$. Then, the interlayer spin-flip tunneling effectively exists.

Now, we consider the interlayer magnetoresistance. Applying the mean field approximation, the interlayer Coulomb interaction term describes the interlayer spin-flip tunneling. We calculate the interlayer conductivity caused by this spin-flip tunneling. For simplicity, we assume the strength of the interlayer Coulomb interaction $V$ is a constant and the effect of the mean field $\chi_{\uparrow\downarrow}$ is already included in it. Thus, the effective interlayer hopping is given by

$$-\eta \sum_{l,n,k,\sigma} \left( c^\dagger_{l,n,k,\sigma} c_{l+1,n,k,-\sigma} + h.c. \right), \tag{9}$$

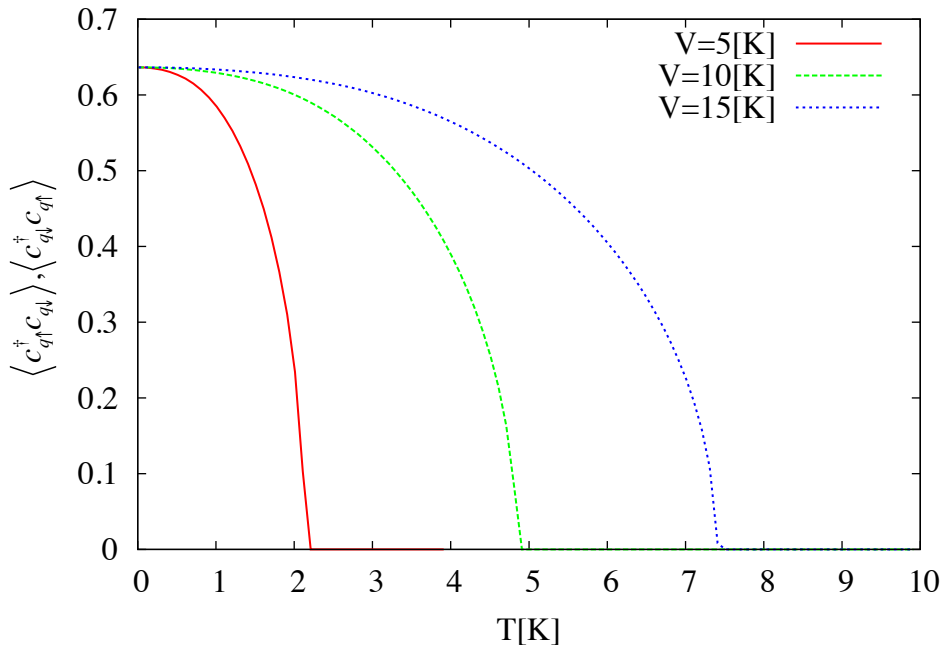

Fig. 1(Color online) The temperature dependence of $\chi_{\uparrow\downarrow}$. The critical temperature increases with increasing the strength of the interlayer Coulomb potential.

where $\eta$ is the vertical interlayer hopping parameter. This describes the vertical interlayer tunneling. However, since the non-vertical interlayer tunneling leads to the inter-Landau level mixing, we need to include the effect of the inter-Landau level mixture in the calculation of the current operator. The current operator is derived from the continuity equation. We calculate the current operator using eq. (9) and the result is

$$\hat{J} = ia_z \eta_{c'} \sum_{l,n,n',k,k'\sigma} \left( c^\dagger_{l,n,k,\sigma} c_{l+1,n',k',-\sigma} + h.c. \right), \tag{10}$$

where $\eta_{c'}$ is the non-vertical interlayer hopping parameter.

We calculate the interlayer conductivity using the Kubo formula,

$$\sigma^{(1)}_{zz} = -\frac{i}{S} \sum_{\alpha,\beta} \frac{f(E_\alpha) - f(E_\beta)}{E_\alpha - E_\beta} \frac{\left|\langle \alpha | \hat{J} | \beta \rangle\right|^2}{E_\alpha - E_\beta + i\delta}. \tag{11}$$

Here, $f$ is the Fermi distribution function and $S$ is the area of the conduction layers. $\alpha$ and $\beta$ represent the single body quantum states. We introduce the matrix elements for non-tilted case $M_{n,n'}$ derived by Morinari and Tohyama[12]. The strength of the inter-Landau level mixture is characterized by $c'_y$. Although the matrix elements are defined for the interlayer transfer integral, we replace it to $\eta_{c'}$. Thus, the interlayer conductivity is given by

$$\sigma^{(1)}_{zz} = \frac{1}{l_B^2 S} \sum_{n,n',\sigma} |M_{n,n'}|^2 \left( -\frac{\partial f}{\partial E} \right)_{E=E_{n,\sigma}} \delta\left( E_{n,\sigma} - E_{n',-\sigma} \right), \tag{12}$$

where $E_{n,\sigma} = \text{sgn}(n) C \sqrt{|n| B} - \sigma g \mu_B B / 2$ is the eigen-energy of the $n$-th Landau level with spin $\sigma$. Hereafter, we assume that $g = 2$ because the spin-orbit coupling is negligible in α-(BEDT-TTF)$_2$I$_3$. From eq. (12), the conductivity of the non-vertical interlayer tunneling without spin-flip is obtained by replacing $\eta_{c'}$ to $t_{c'}$ in the matrix elements and $-\sigma$ to $\sigma$ in the argument of the delta function.

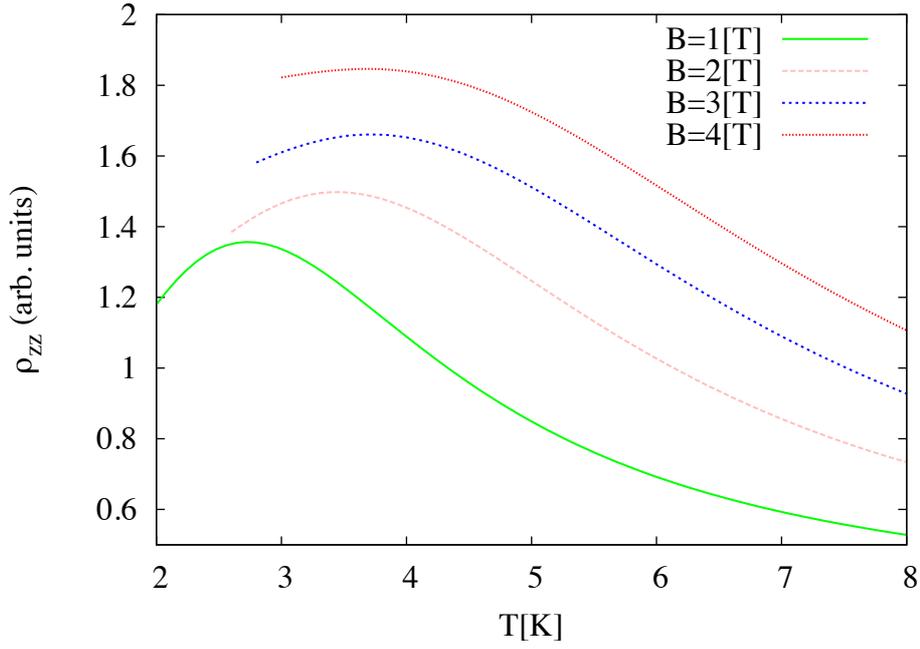

Fig. 2 (Color online) The temperature dependence of the interlayer magnetoresistance with $C = 10 \text{ KT}^{-1/2}$, $\Gamma = 5 \text{ K}$ and $V = 5 \text{ K}$. The peak temperatures increase with increasing the magnetic fields.

In order to include the effect of the Landau level broadening, we replace the delta function with the Lolentzian function with the half value width of $\Gamma$. Assuming $\Gamma$ is a constant, we numerically calculate the temperature dependence of the interlayer magnetoresistance. The result for the parameters $C = 10 \text{ KT}^{-1/2}, \Gamma = 5 \text{ K}$ and $V = 5 \text{ K}$ is shown in Fig. 2. The peak temperatures increase as we increase the magnetic field. The differentiation of the Fermi distribution function has a peak, the width of which broadens by increasing temperature, and so, the overlap with the Lorentzian function increases. For the interlayer spin-flip tunneling, the Lorentzian function is shifted by the Zeeman energy. Consequently, the peak temperature of the magnetic field is shifted by the Zeeman energy.

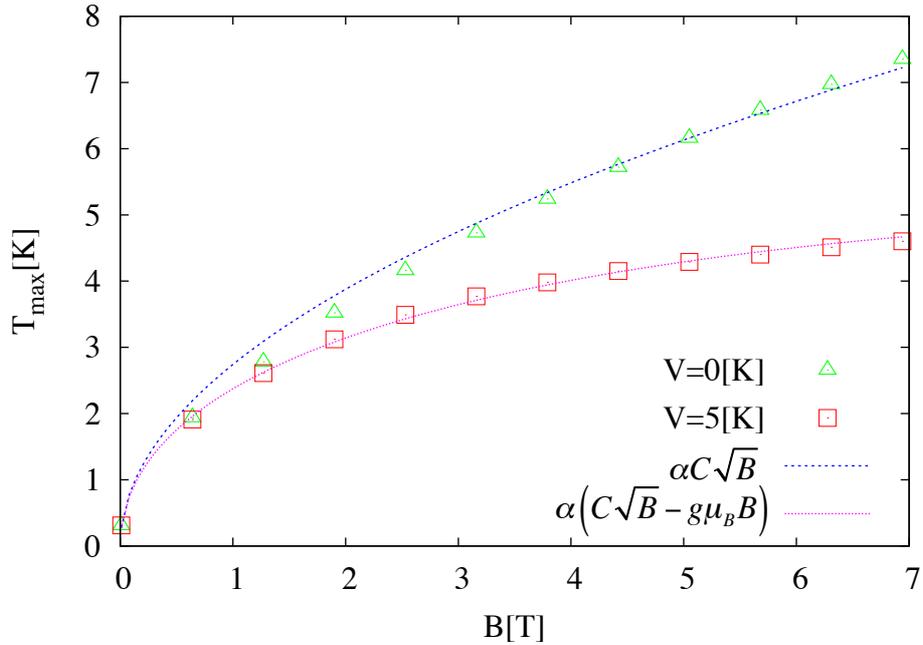

Fig. 3 (Color online) The square points shows the magnetic field dependence of the peak temperature with $C = 10 \text{ KT}^{-1/2}$ and $V = 5 \text{ K}$. $\alpha$ is the fitting parameter of $T_{max}$ with $V = 5 \text{ K}$ to $C\sqrt{B} - g\mu_B B$. The triangle points are the cases without the spin-flip processes for the comparison.

We now calculate the peak temperature for different magnetic fields. The result with $C = 10 \text{ KT}^{-1/2}$ and $V = 5 \text{ K}$ is shown in Fig. 3. This is fitted well to the functional form of $\alpha \left( C\sqrt{B} - g\mu_B B \right)$ with the constant $\alpha \sim 0.27$. Using this value of $\alpha$, the numerical result with $V = 0 \text{ K}$, which corresponds to the case without the interlayer spin-flip tunneling, is fitted well to the functional form of $\alpha C \sqrt{B}$. So, we found that the characteristic energy of the interlayer magnetoresistance is shifted by the Zeeman energy when the interlayer spin-flip tunneling exists. The result is valid when $\chi_{\uparrow\downarrow}$ is finite. $\chi_{\uparrow\downarrow}$ vanishes above the critical temperature as shown in Fig. 1. However, assuming $\chi_{\uparrow\downarrow}$ as a constant, we obtain the peak temperature in quantitatively good agreement with the experimental data in ref. 13 except the coefficient. The mean field $\chi_{\uparrow\downarrow}$ is finite below a characteristic temperature. This temperature is determined by the interlayer Coulomb interaction. In this system, the number of conduction electrons is very small. So, the screening effect is negligible[17]. Therefore, the interlayer Coulomb interaction is on the order of 20K. The coefficient $\alpha$ is about one third of the experimentally evaluated value. This would be because the magnetic field or temperature dependence of $\Gamma$ are not considered here. The overlap of the Lorentzian function and the differentiation of the Fermi distribution function can be evaluated more precisely. This is a subject for a future study.

To conclude, we have examined the effect of the interlayer spin-flip tunneling. The opposite spin mean field arising from the interlayer Coulomb interaction leads to spin-flip tunneling. The magnetic field dependence of the peak temperature of the interlayer magnetoresistance is shifted by the Zeeman energy and is in good agreement with the experiment.

**Acknowledgment**

This work was financially supported in part by a Grant-in-Aid for Scientific Research (A) on "Dirac Electrons in Solids" (No. 24244053) and a Grant-in-Aid for Scientific Research (B) (No. 25287089) and (C) (No. 24540370) from the Ministry of Education, Culture, Sports, Science and Technology, Japan.

*kubo.kenji.22x@st.kyoto-u.ac.jp